\documentclass{moriond}

\usepackage{amsmath}

\def\Journal#1#2#3#4{#4, {#1}, {\bf #2}:#3}

\def\mnras{\em MNRAS}
\def\apj{\em ApJ}
\def\nat{\em Nature}
\def\aapr{\em A\&A Review}
\def\aap{\em A\&A}

\def\prd{\em Physical Review D}

\def\be{\begin{equation}}
\def\ee{\end{equation}}
\def\bea{\begin{eqnarray}}
\def\eea{\end{eqnarray}}

\newcommand{\Photo}{}

\begin{document}
\vspace*{4cm}
\title{PARTICLE ACCELERATION IN GALACTIC WIND BUBBLES}

\author{Enrico Peretti}

\address{Niels Bohr International Academy, Niels Bohr Institute, University of Copenhagen, \\ Blegdamsvej 17, DK-2100, Copenhagen, Denmark}

\maketitle\abstracts{
Winds are ubiquitous in galaxies and often feature bubble structures. 
These wind bubbles are characterized by
an external forward shock expanding in the surrounding medium and a wind termination shock
separating the cool and fast wind from the hot shocked wind. 
While the forward shock could not be able to accelerate particles efficiently for a long time, at the wind termination shock the necessary
conditions for efficient acceleration may be present.
We develop a model for particle acceleration at the termination shock of such bubbles analysing
the consequences of different possible engines powering the wind. 
We finally explore the multi-messenger potential of galactic winds in terms of escaping cosmic rays and high-energy gamma rays and neutrinos produced through hadronic interactions.}

\section{Introduction}

In the last decades cosmic rays (CRs) have been observed from sub-relativistic energies up to $\sim 10^{20}$ eV. While there is a general consensus on the Galactic origin of CRs with energies below the \textit{Knee}, where $E_{\rm Knee} \simeq 3 \times 10^{15}$ eV, the nature of the astrophysical phenomena responsible for particle acceleration to higher energies as well as their location in the Universe are still unanswered questions \cite{Blasi-review2013}. 

Galactic winds (GWs) are one of the most promising extragalactic candidates for accelerating energetic particles beyond $E_{\rm Knee}$ as they possess a large power content ($10^{39-46} \, \rm erg \, s^{-1}$) and typical lifetimes exceeding several Myr; a time sufficiently longer than typical timescales of CR acceleration. 
This large amount of power available for such a long time can be partially converted into populations of relativistic particles injected through diffusive shock acceleration (DSA) at the strong shocks developed by these systems. 
GWs are also very common in galaxies as they can be powered by several different engines such as active galactic nuclei (AGNi), starburst nuclear regions (SBNi) and the pressure gradient of CRs leaving galactic disks \cite{Veilleux,Zhang2018}. 
While CR-driven winds \cite{Breitshwerdt91,Recchia2016} may not reach the necessary conditions for an efficient particle acceleration above the PeV range, the outflows powered by AGNi \cite{King,Fauchere2012,Fiore17,Menci:2019aih} and SBNi \cite{Strickland-Stevens-2000,Elizabete_2013,Owen1,Owen2} can reach very high terminal velocity, $V_{\infty}=10^3-10^4 \, \rm km \, s^{-1}$, and mass-loss rate, $\Dot{M} = 0.1-10^3 \, \rm M_{\odot} \, yr^{-1}$, allowing them to accelerate particles up to $10^2 \, \rm PeV$ and possibly beyond \cite{Peretti-wind}.

The presence of highly energetic particles and associated non-thermal photon and neutrino emission has been explored both in AGN-driven \cite{Lamastra1,Liu2018,Lamastra2} and starburst-driven outflows \cite{Romero-Ana,Muller-Romero}. 
However, a detailed understanding of the acceleration mechanisms and transport conditions of high-energy (HE) particles in these systems has not been achieved yet. 

Following the model developed in MBPC21 \cite{MBPC2021} and PMBC22 \cite{Peretti-wind}, in this work we present a model for particle acceleration at the wind termination shock of galactic winds featuring a bubble structure and discuss their multimessenger implications in terms of CRs protons above the Knee, gamma rays and HE neutrinos, hereafter labelled as ($p$,$\gamma$,$\nu$) . 

\section{Galactic wind bubble}

The intense activity of SBNi and AGNi can launch and sustain highly mass loaded and fast GWs for several Myr. 
These outflows are enough powerful to quickly break out galactic disks and expand in the galactic halos where they feature a bubble structure.
Independent of the nature of the engine, the wind bubble has a standard evolution provided that the engine activity remains approximately stationary and there are not sudden changes in the properties of the external medium \cite{Weaver77,Koo-McKee,Koo-McKee2}.

\begin{figure}[h]
    \centering
	\includegraphics[width=0.65\linewidth]{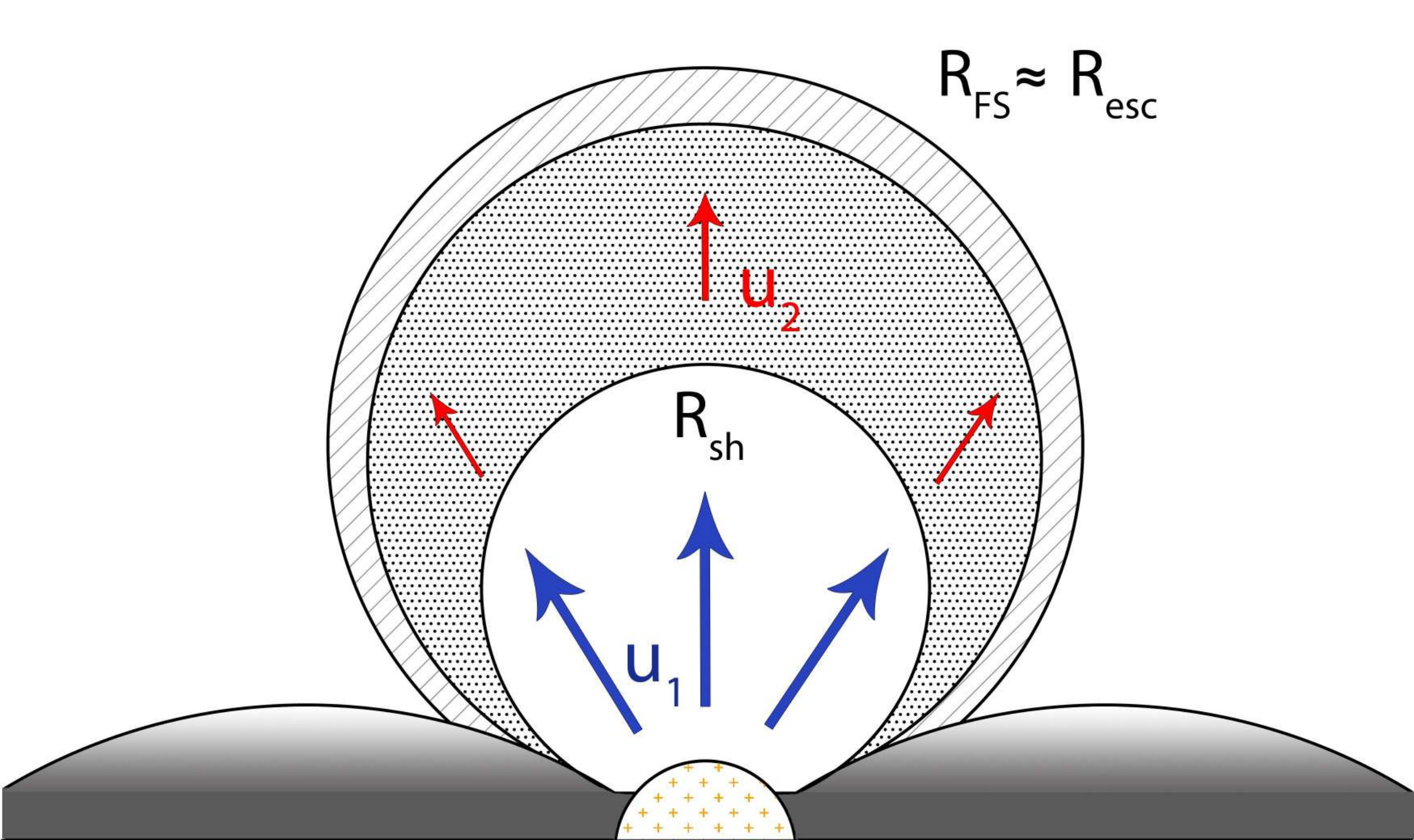}
\caption[]{Structure of a lobe of a galactic wind bubble where the blue (red) arrows represent the fast cool (hot shocked) wind region. $R_{\rm sh}$ ($R_{\rm fs}$) identifies the location of the wind shock (forward shock).}
\label{Fig: wind-bubble}
\end{figure}

As a an SBN or an AGN enters an active phase a supersonic outflow can be launched.
Since the wind is supersonic, a forward shock forms ahead of the outflow, while the collision of the wind material with the external medium leads to the formation of a second shock projected towards the central engine inside the ejected material, the wind termination shock (hereafter \textit{wind shock}). 
The two shocks are separated by a contact discontinuity, the physical boundary of the wind material. 
A sketch of the structure of a galactic wind bubble is reported in Fig.~\ref{Fig: wind-bubble}, where the blue (red) lines represent the fast cool (hot shocked) wind region which is also labelled by the subscript 1 (2). Hereafter region 1 (2) will be also referred as upstream (downstream) region. 
The upstream region is separated from the downstream region by the wind shock ($R_{\rm sh}$), while the contact discontinuity and the forward shock ($R_{\rm fs}$), which are located very close to each other, set the boundary between the bubble and the halo.

The first phase of the outflow evolution is characterized by a free expansion of the wind material with the two shocks expanding close to each other with constant velocity.
After a time $t_{\rm dec}$, when the mass of swept-up halo material accumulated between the forward shock and the contact discontinuity becomes comparable with the total mass in the wind, the free expansion terminates and the outflow decelerates. 
The deceleration phase is characterized by a self-similar evolution of the shock radii where the wind shock and the forward shock evolve respectively as $R_{\rm sh} \sim t^{2/5}$ and $R_{\rm fs} \sim t^{3/5}$. 
By the time the forward shock keeps expanding with low Mach number, the wind shock decelerates faster while remaining strong, thereby creating ideal conditions for the DSA process in stationary conditions. 
In addition, the different deceleration characterizing the two shock locations makes the hot bubble (the shocked wind region) expanding with time. 
At later time the pressure in the hot bubble drops down to the level of the undisturbed halo medium and the bubble enters a pressure confined stage in which the wind shock is practically stalled at the location where the free wind ram pressure balances the one in external medium:
\begin{equation}
    \label{Eq: Shock-radius}
    R_{\rm sh} = \sqrt{\Dot{M} V_{\infty}/[4 \pi P_h]} \, ,
\end{equation}
where $P_h$ is the pressure outside the wind bubble.

Once the plasma leaves the launching region it quickly approaches the terminal velocity \cite{ChevalierClegg85}, $V_{\infty}$, so that with good approximation one can assume a spatially constant wind speed $u_1 \approx V_{\infty}$ in the whole upstream region ($r<R_{\rm sh}$); consequently $\rho_1(r) \simeq \Dot{M}/[4 \pi r^2 u_1]$.
The downstream region is approximately adiabatic therefore it is characterized by constant density, $\rho_2 = \mathcal{R} \rho_1$, and a velocity profile $u_2(r) = u_2 [R_{\rm sh}/r]^{2}$, where $u_2 = u_1/\mathcal{R}$ and $\mathcal{R}$ is the compression ratio at the shock ($\mathcal{R}=4$ for strong shocks).

The magnetic field is assumed to be of turbulent nature with a coherence length $l_c \sim 10^2 \, \rm pc$, while the magnetic field pressure $U_{B}$ is estimated as a fraction $\epsilon_B = 10\%$ of the kinetic energy density of the plasma in the whole fast cool wind region: 
\begin{equation}
\label{Eq: Bfield}
    B_1(r) = [\epsilon_B \Dot{M} u_1]^{1/2} \, r^{-1} \, .
\end{equation}
At the wind shock the magnetic field gets compressed of the standard factor $\sqrt{11}$ and stays constant throughout the shocked region. 

\section{Model}

The phase space density of cosmic rays in the wind bubble is computed by solving the stationary CR transport equation in spherical symmetry 
\begin{equation}
    \label{Eq: transport}
    r^2 u(r) \partial_r f(r,p) = \partial_r [D(r,p) \partial_r f(r,p)] + \frac{1}{3} p \, \partial_p f(r,p) \, \partial_r[r^2 u(r)] + Q(r,p) + L(r,p) \, ,
\end{equation}
where $f$ is the cosmic ray phase space density, $D$ is the diffusion coefficient, $Q$ is the injection term, $L = n \sigma_{pp} c f$ is the energy loss term accounting for pp interactions and $c$ is the speed of light. 
The diffusion coefficient is computed in the framework of quasi-linear theory as
\begin{equation}
    \label{Eq: Diff_coeff}
    D(r,p) = 
    \begin{cases}
    \frac{1}{3} v(p) \, r_{\rm L}(r,p)^{2-\delta} \, l_c^{\delta - 1} & r_{\rm L} \leq l_c \\ 
    \frac{1}{3} v(p) \, l_c \, \Big[\frac{r_{\rm L}(r,p)}{l_c} \Big]^{2} & r_{\rm L} > l_c
    \end{cases}
    \, ,
\end{equation}
where $v(p)$ is the particle velocity, $r_{\rm L}$ is the Larmor radius and $\delta$ is the power law slope of the turbulence spectrum. We assume $\delta=3/2$ as prescribed for a magnetohydrodynamical (Kraichnan-like) turbulence cascade.
The injection term $Q$ embeds the physics of the DSA and reads:
\begin{equation}
    \label{Eq: injection}
    Q(r,p) = \frac{\eta \, n_1 \, u_1}{4 \pi p^2} \, \delta[p-p_{\rm inj}] \, \delta[r-R_{\rm sh}] \, ,
\end{equation}
where $p_{\rm inj}$ is the injection momentum, $n_1$ is the plasma density at the shock and $\eta$ is the efficiency parameter regulating the number of wind particles entering the DSA process. 
In particular, $\eta$ is taken such that the CR pressure at the shock is $10\%$ of the wind ram pressure and $p_{\rm inj}$ does not play any relevant role in the overall normalization.

Eq.~\eqref{Eq: transport} is solved by integrating separately the upstream region and the downstream region and finally joining the two solutions at the shock location following the approach described in MBPC21 and PMBC22. 
Given the spherical symmetry of the problem a 0-flux condition is assumed at the center of the system, namely $[D\partial_r f- uf]_{r=0}=0$.
Two possible scenarios are considered for particles reaching the edge of the bubble: A) flux conservation at the forward shock, $[D\partial_r f- uf]_{r=R_{\rm fs}}=D_h\partial_r f|_{\rm R_{r =\rm fs}}$, and diffusive transport in the halo medium characterized by a diffusion coefficient $D_h(p)$; B) free escape boundary conditions at the forward shock, $f_{\rm fs}=0$ and $j_{\rm esc} \neq 0$, and ballistic propagation in the halo. 
In both scenarios, the phase space density of CR and the associated flux are negligible at large distance from the system. 

The interested reader is referred to PMBC22 for details of the analytical derivation, the iterative technique and the calculation of the gamma rays and neutrinos, whereas the following discussion will focus on the results and their physical implications.
The solution in the upstream region reads
\begin{equation}
    \label{Eq: up} 
    f_1^k(r,p) = f_{\rm sh}^k(p) \, e^{-\int_r^{R_{\rm sh}} dr' \, V_{\rm eff}^k(r',p)/D_1(r',p)} \, ,
\end{equation}
where $k= \rm A, B$ is the index representing the scenario, $f_{\rm sh}^k(p)$ is the solution at the shock and $V_{\rm eff}^k$ is the effective velocity experienced by particles in the upstream region. Its analytical expression is the following:
\begin{gather}
    \label{Eq: V_eff}
    V_{\rm eff}^k(r,p) = u_1 + [G_1^k(r,p)+H_1^k(r,p)] / [r^2 f_1^k(r,p)] \\
    \label{Eq: G_up}
    G_1^k(r,p) = \frac{1}{3} \int_0^{r} dr' \, \partial_{r'}[r'^2 u_1] \, f_1^k(r',p) \, \Big[- \frac{\partial {\rm ln} p^3f_1^k(r',p) }{ \partial {\rm ln} p } \Big] \\
    \label{Eq: H_up}
    H_1^k(r,p) = \int_0^{r} dr' \, r'^2 \, f_1^k(r',p) \, n(r') \sigma_{pp}(p) c
\end{gather}
As in the standard case of the infinite planar shock \cite{Blasi-review2013}, the transport of particles in the upstream region is regulated by the competition between diffusion, which tries to spatially isotropize particles, and advection, which pushes the same particles back to the shock. 
This competition sets the exponential behavior of Eq.~\eqref{Eq: up} where, different from the planar case, the effective velocity $V_{\rm eff}$ embeds also information on the geometry of the system and energy losses.
The solution in the downstream region is characterized by negligible energy losses and, different from the upstream region, it has an analytic form which differs among the two considered scenarios. In particular, in scenario (A) the solution reads
\begin{equation}
    \label{Eq: down-Diff-halo}
    f_2^{\rm A}(r,p) = f_{\rm sh}^{\rm A}(p)  \, e^{\alpha(r,p)} \frac{1+\beta(p)[e^{\alpha(R_{\rm fs},p)-\alpha(r,p)}-1]}{1+\beta(p)[e^{\alpha(R_{\rm fs},p)}-1]} \, ,
\end{equation}
where $\alpha(r,p)= R_{\rm sh} u_2 (1 - R_{\rm sh}/r)/D_2(p)$ and $\beta(p) = R_{\rm fs} D_{\rm h}(p)/[u_2 R_{\rm sh}^2]$.
The downstream solution in scenario (B) is the same as derived in MBPC21:
\begin{equation}
    \label{Eq: down-Ball-halo}
    f_2^{\rm B}(r,p) = f_{\rm sh}^{\rm B}(p)  \, \frac{[1-e^{\alpha(r,p)-\alpha(R_{\rm fs},p)}]}{[1-e^{-\alpha(R_{\rm fs},p)}]} \, .
\end{equation}
\begin{figure}[h]
    \centering
	\includegraphics[width=0.55\linewidth]{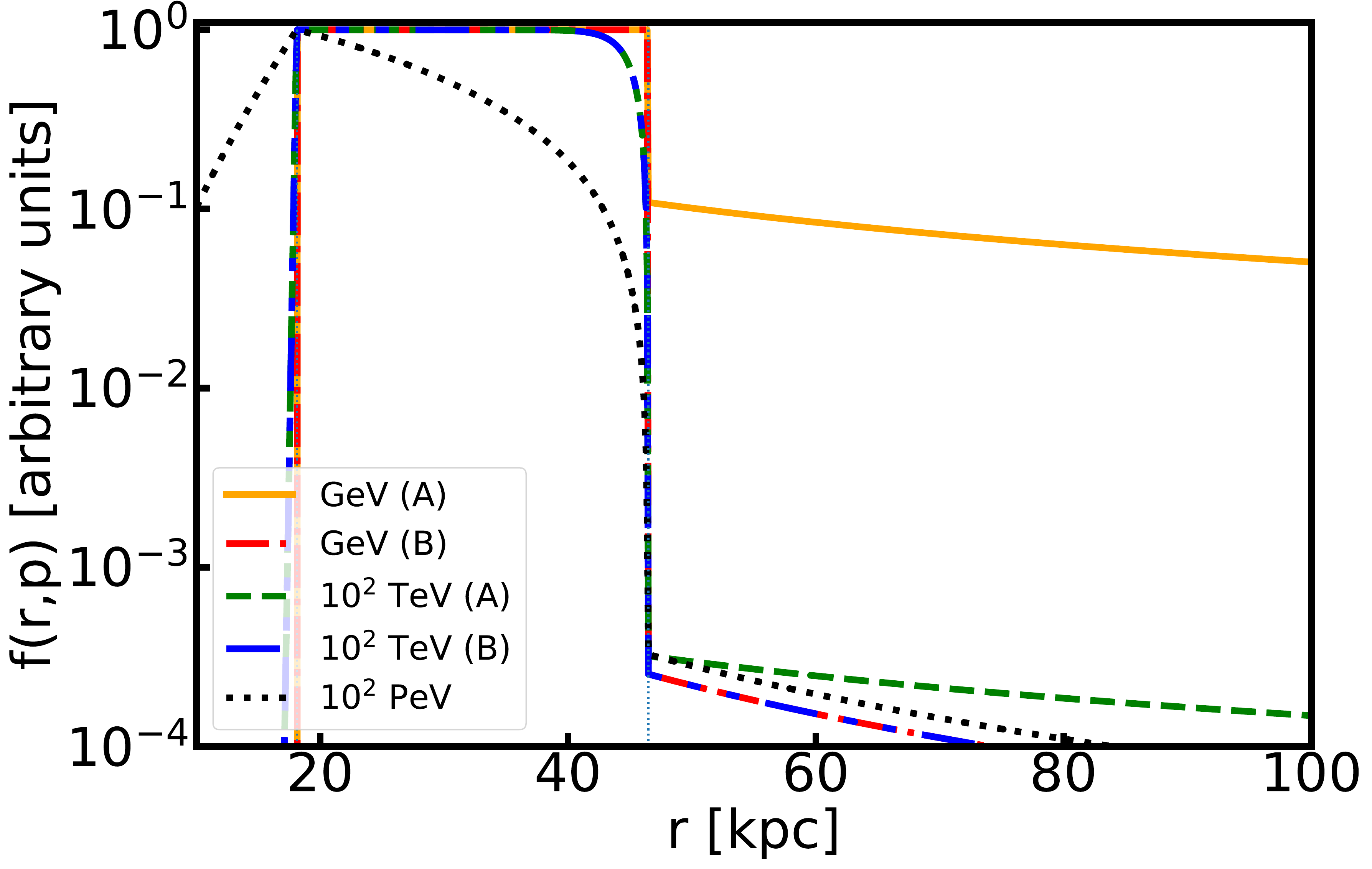}
\caption[]{Radial behavior of the CR phase space density $f$ normalized to its value at the shock $f_{\rm sh}$, where $R_{\rm sh}\approx 18 \, \rm kpc$ and $R_{\rm fs}\approx 46 \, \rm kpc$. Scenario A and B are compared at different energies. In order: Thick yellow line (A,1 GeV) and red dot-dashed (B,1 GeV); green dashed line (A,$10^2$ TeV) and blue dot-dot-dashed line (A,$10^2$ TeV). The two scenarios do not show relevant difference at the maximum energy (black dotted line). }
\label{Fig: Particles-radial}
\end{figure}

The CR phase space density in the halo for the two scenarios can be written as
\begin{gather}
    \label{Eq: halo-d}
    f_{3}^{\rm A}(r,p) = f_{\rm sh}^{\rm A}(p) \frac{e^{\alpha(R_{\rm fs},p)}}{1 + \beta(p) (e^{\alpha(R_{\rm fs},p)}-1)} \, \frac{R_{\rm fs}}{r} \\
    \label{Eq: halo-b} 
    f_{3}^{\rm B}(r,p) = f_{\rm sh}^{\rm B}(p) \, \frac{u_2/c}{1-e^{-\alpha(R_{\rm fs},p)}} \Big( \frac{R_{\rm sh}}{r} \Big)^2 
\end{gather}
where one can identify the different transport regimes for the solution in the two scenarios, namely the $1/r$ dependence for the diffusion dominated scenario (A) and the ballistic behavior typical of scenario (B).

Fig.~\ref{Fig: Particles-radial} illustrates the radial behavior of the solution in the whole bubble and halo under the assumption of $\Dot{M}=10 \rm \, M_{\odot} \, yr^{-1}$ and $V_{\infty}= 3000 \, \rm km \, s^{-1}$, where an $E_{\rm max}\simeq 10^2 \, \rm PeV$ is obtained. 
The external diffusion coefficient $D_h$ is computed assuming $\delta=3/2$ and with $B_h \approx 10^{-10} \, \rm G$, $l_{\rm c,h} \sim 10^2 \, \rm Mpc$ as inferred from UHECR hotspots \cite{Neronov_D_h}. 
The two scenarios (A,B) are compared at different energies, 1 Gev and $10^2$ TeV, whereas at the maximum energy there is no appreciable difference. 
In particular, one can see that while the free escape boundary condition (scenario B) equally suppresses the CR phase space density in the halo, the diffusive halo scenario (A) is characterized by an energy dependent suppression and consequently a possible high density of sub TeV particles in the halo. Given the large value of $D_h$ scenario A and B are practically indistinguishable in the downstream region. Finally, in the upstream region only particles with energy larger than $\sim \rm PeV$ are not confined in the immediate shock vicinity by the wind and there is no appreciable difference between the two scenarios.
\begin{figure}[h]
    \centering
	\includegraphics[width=0.48\linewidth]{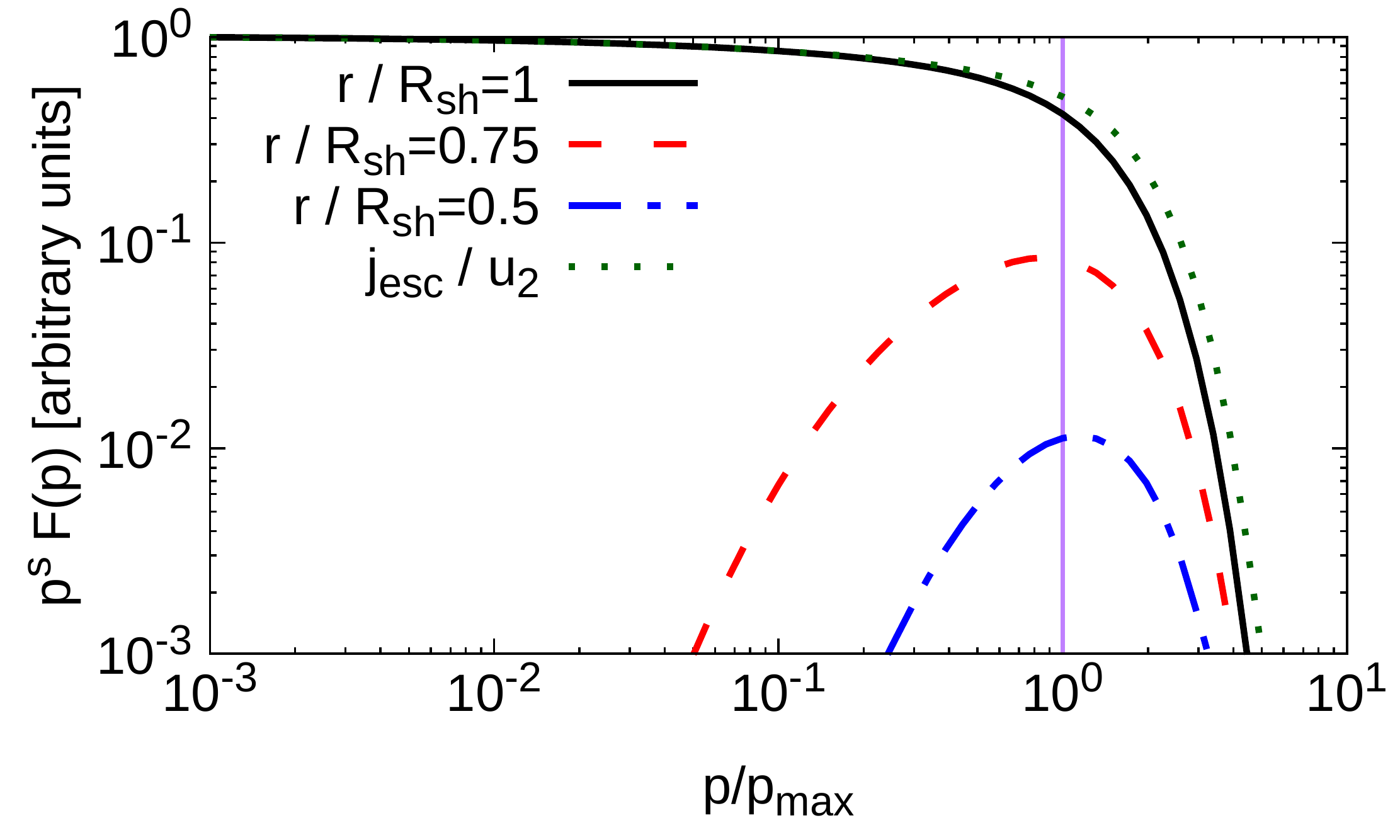} \quad \includegraphics[width=0.48\linewidth]{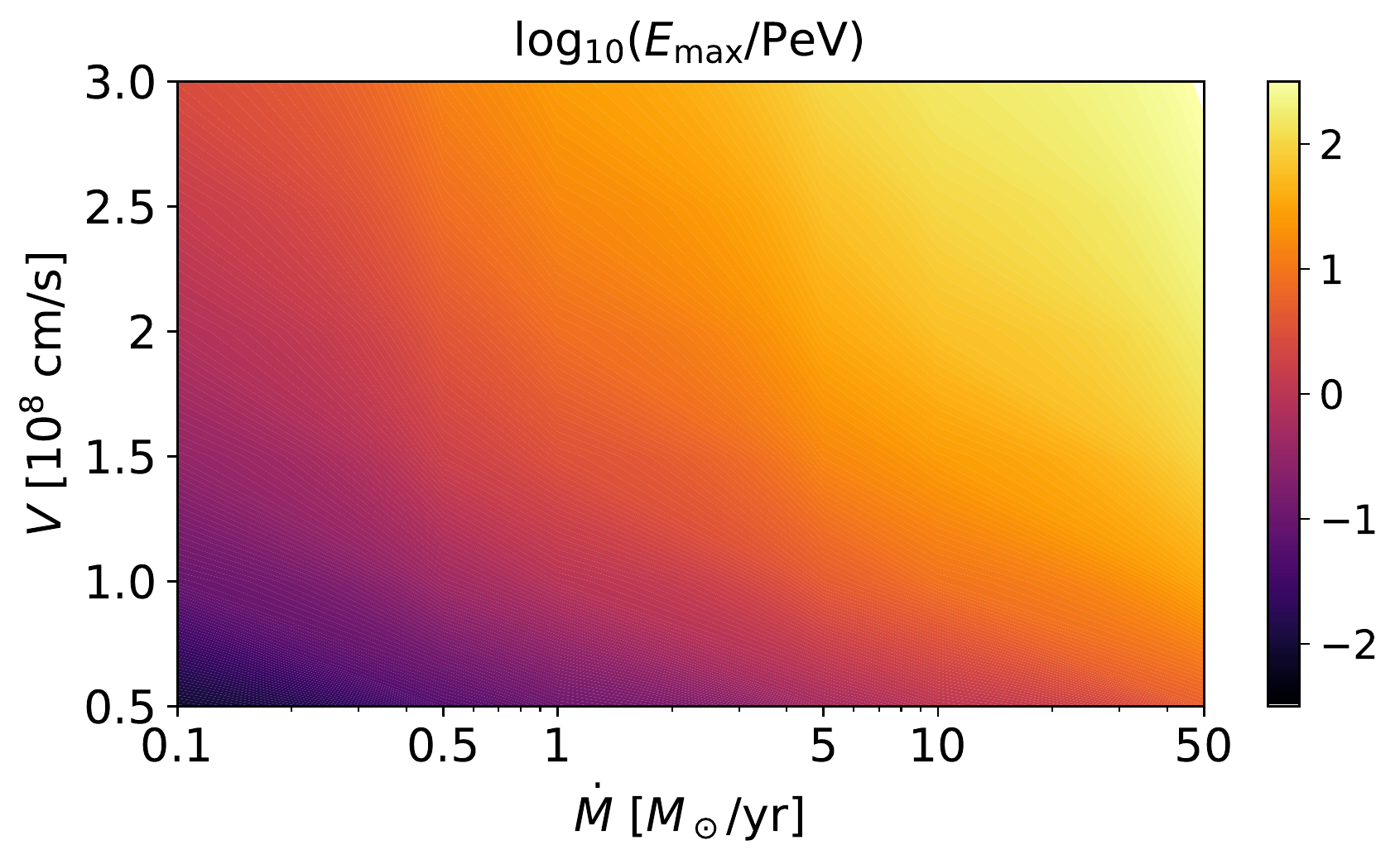}
\caption[]{Left Panel: Particle spectra at the wind shock (thick black line) compared with the solution at 75\% (red dashed line) and 50\% (blue dot-dashed line) of the shock radius in the upstream. The escaping flux for scenario B (green dotted line) is also shown. Right panel: Maximum energy as a function of the main macroscopic properties of the wind ($\Dot{M}$,$V_{\infty}$).}
\label{Fig: Particles}
\end{figure}

The solution at the shock reads
\begin{equation}
    \label{Eq: Shock}
    f_{\rm sh}^k(p) = \frac{s n_1 \eta}{4 \pi p_{\rm inj}^3} \Big( \frac{p_{\rm inj}}{p} \Big)^s {\rm exp}[-\Gamma_1^k(p)] \, {\rm exp}[-\Gamma_2^k(p)] \, ,
\end{equation}
where $\Gamma_{1(2)}^k$ is the cut off function associated to the upstream (downstream) which has the following expression: 
\begin{gather}
    \label{Eq: Gamma_1}
    \Gamma^k_1(p) = s \int_{p_{\rm inj}}^p \frac{dp'}{p'} \frac{G_1^k(R_{\rm sh},p')+H_1^k(R_{\rm sh},p')}{u_1 R_{\rm sh}^2 f_{\rm sh}^k(p')} \\
    \label{Eq: Gamma_2_A}
    \Gamma^{\rm A}_2(p) = s \int_{p_{\rm inj}}^p \frac{dp'}{p'} \frac{(u_2/u_1)[1-1/\beta(p')]}{ {\rm exp}[\alpha(R_{\rm sh},p')]-1 + 1/\beta(p') } \\
    \label{Eq: Gamma_2_B}
    \Gamma^{\rm B}_2(p) = s \int_{p_{\rm inj}}^p \frac{dp'}{p'} \frac{u_2/u_1}{{\rm exp}[\alpha(R_{\rm sh},p')]-1}
\end{gather}

Fig.~\ref{Fig: Particles} shows the particle spectrum at different radii in the bubble (left panel) and the dependence of $E_{\rm max}$ (right panel) on the main macroscopic properties of the wind, namely $\Dot{M}$ and $V_{\infty}$.
As expected from the standard DSA, the particle spectrum at the shock (black line) features a $p^{-s}$ behavior up to the maximum momentum $p_{\rm max} \simeq 10^2 \, \rm PeV/c$ while innermost solutions (red dashed and blue dot-dashed) are characterized by a low energy cut off in agreement with the radial suppression shown in Fig.~\ref{Fig: Particles-radial}. The escaping flux (green dotted line) obtained in scenario (B) is also shown and its spectral shape is very similar to the shock solution.
In the right panel it is possible to observe that, for standard parametric assumptions, both wind bubbles can achieve maximum energy as large as $\sim 10^2$ PeV. This suggests that in scenarios with higher wind speed velocity the EeV range could be unlocked. 

It is interesting to notice that Eq.~\eqref{Eq: down-Ball-halo} as well as Eq.~\eqref{Eq: Gamma_2_B} can be obtained respectively as limits of Eq.~\eqref{Eq: down-Diff-halo} and \eqref{Eq: Gamma_2_A} for $\beta \rightarrow \infty$, namely when particles do not diffuse anymore but propagate ballistically in the halo.
This explains the analytic difference between current results and what previously obtained \cite{Morlino_ICRC} on the particle transport outside star cluster wind bubbles, where the geometry of the problem is very similar.
\begin{figure}[h]
    \centering
	\includegraphics[width=0.40\linewidth]{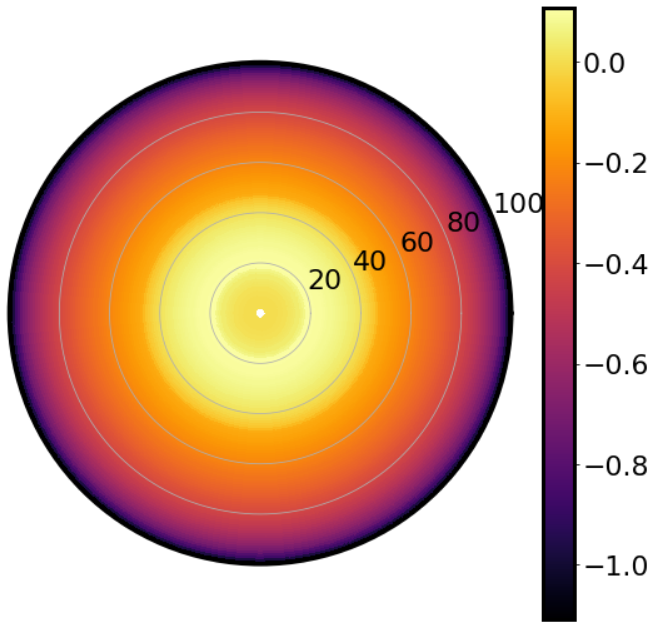} \quad \includegraphics[width=0.40\linewidth]{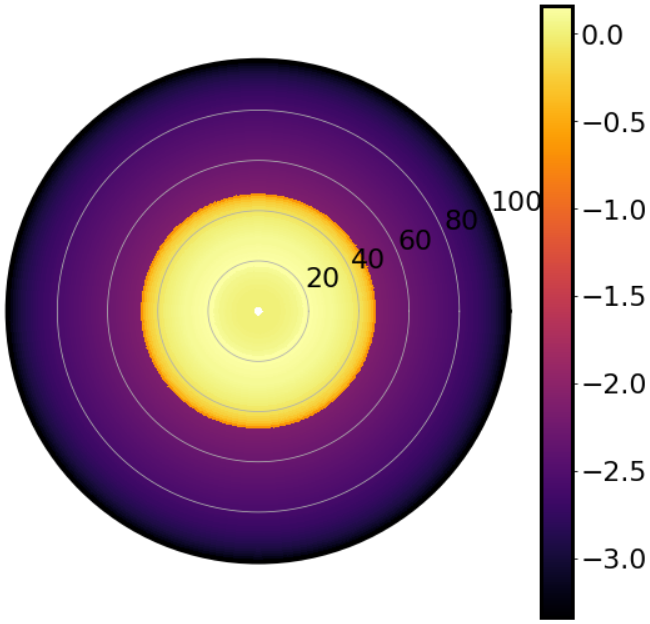}
\caption[]{Gamma-ray sky-map-projections at 1 GeV for scenario A (left panel) and scenario B (right panel). The color code is obtained by re-scaling the luminosity to the one at the innermost radius. The grid numbers indicate the distance in kpc from the central engine. The geometrical center of the system where the engine is present is masked and shown as white dot.}
\label{Fig: Halo-map}
\end{figure}
The transport regime in the galactic halo impacts the way halos shine in gamma. 
Fig.~\ref{Fig: Halo-map} shows the gamma-ray sky-map-projections associated to Fig.~\ref{Fig: Particles-radial} for scenario A (left panel) and scenario B (right panel) computed assuming a target density $n_h = 5 \cdot 10^{-3} \, \rm cm^{-3}$ in the halo. 
The flux is computed at 1 GeV and it is obtained performing a line-of-sight integration of the emissivity. 
Finally, the color code highlights the emissivity relative to the innermost line of sight, whereas the geometrical center of the system has been masked with a white spot. %
By comparing the two scenarios one can conclude that, despite the wind bubble emission dominates, galactic halos where CRs are actually diffusing can be up to two orders of magnitude brighter than halos where CRs escape balistically.

A comment on the transport in the halo is in order: in the results shown above a nominal target halo density $n_h = 5 \cdot 10^{-3} \, \rm cm^{-3}$ has been assumed.
While, clearly, a lower density would result in a lower halo luminosity, a scenario with dense clouds in the halo or nearby satellite galaxies could result in the halo outshining the wind bubble.

\section{Discussion and conclusions}

The DSA can efficiently take place at wind shocks in galactic wind bubbles where CRs can be accelerated up to at least $\sim 10^2$ PeV. 
Such high energy particles in the dense and polluted environment characterizing active and star forming galaxies implies a great (p,$\gamma$,$\nu$) multimessenger potential for the associated wind bubbles.
In addition, given the large number of SBGs and AGNi in the Universe, these sources are also very promising candidate for a multimessenger (p,$\gamma$,$\nu$) diffuse flux.
\begin{figure}[h]
    \centering
    \includegraphics[width=0.47\linewidth]{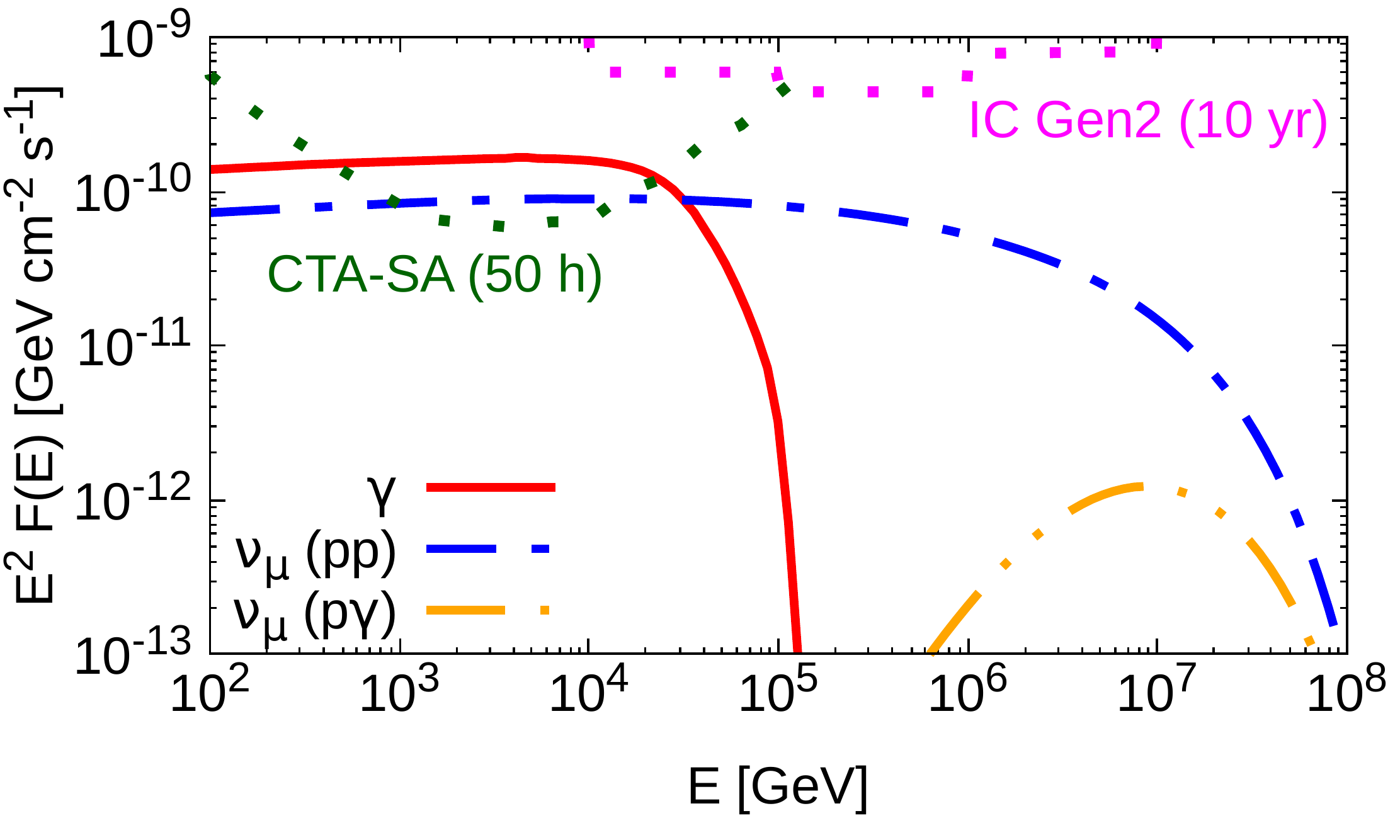} \quad
	\includegraphics[width=0.47\linewidth]{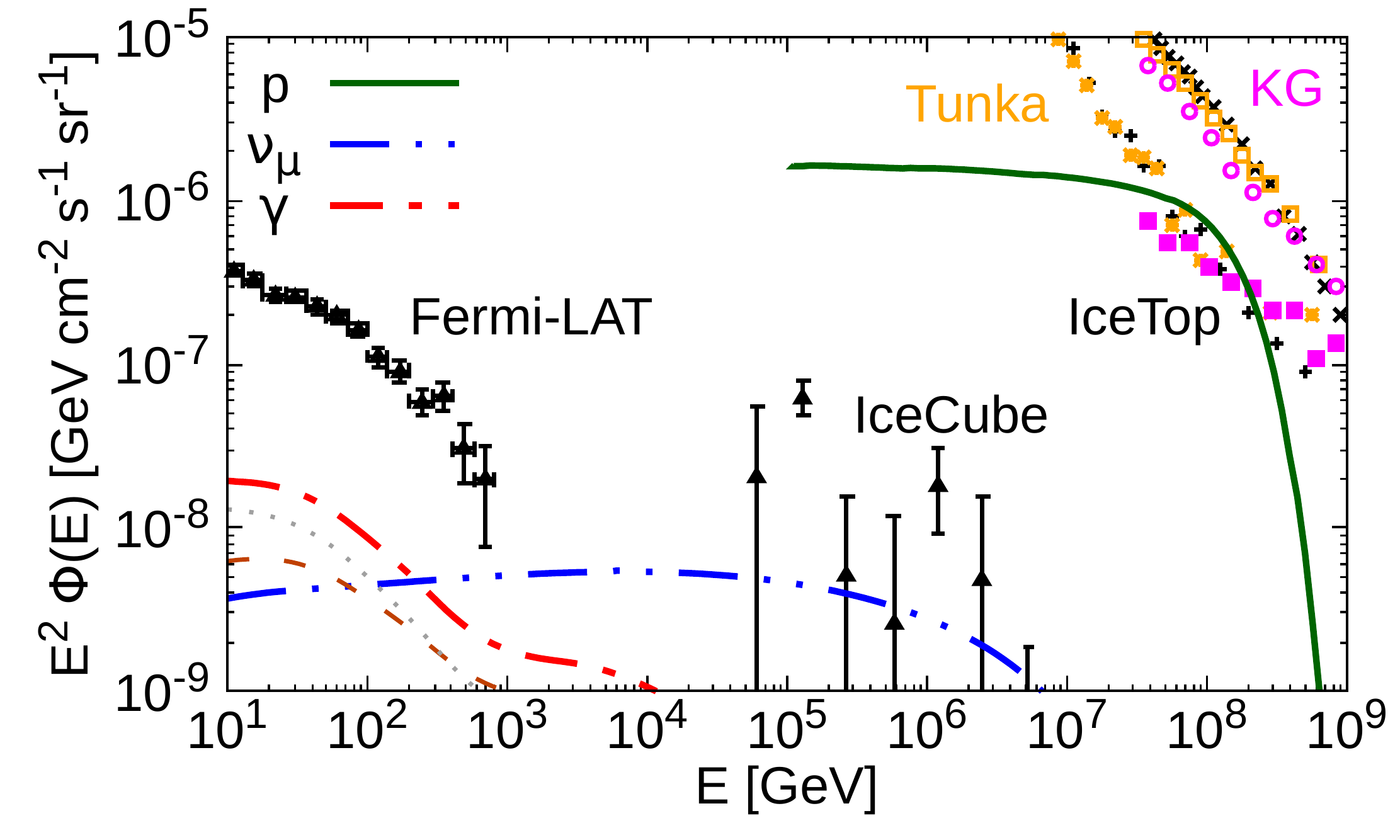} 
\caption[]{Left panel: multimessenger ($\gamma$,$\nu_{\mu}$) spectrum from an isolated wind bubble located at $D_{\rm L}=3.9$ Mpc. Gamma rays (thick red line) and neutrinos produced via pp (blue dot-dashed) and $p \gamma$ (orange dot-dot-dashed) are compared with CTA \cite{CTA} southern array (green dots) and IceCube-Gen2 \cite{IceCube-Gen2} (magenta dots) pointlike sensitivities. 
Right panel: Multimessenger (p,$\gamma$,$\nu_{\mu}$) diffuse flux produced by starburst-driven wind bubbles with line-style (thick green, dot-dashed red, dot-dot-dashed blue). 
The direct gamma rays (brown dashed line) are shown separately from the cascade component (grey dotted line).
The model prediction is compared with the gamma-ray data of Fermi-LAT \cite{Fermi-LAT<820}, IceCube neutrinos \cite{IceCube2020_LAST}, and CRs observed bt Tunka \cite{Tunka2013,Tunka2016} (orange points), Kascade-Grande \cite{Kascade-GRANDE2017} (magenta points) and IceTop \cite{IceTop2019} (black points). For each CR detectors the light component (stars, pluses, filled squares) is shown separately from the associated all-particle spectrum (open squares, crosses, open circles).}
\label{Fig: Multimessenger-Diff}
\end{figure}

The left panel of Fig.~\ref{Fig: Multimessenger-Diff} shows the multimessenger ($\gamma$,$\nu_{\mu}$) flux produced by a wind bubble under the assumption of $\Dot{M} = 10 \, M_{\odot} \, yr^{-1}$, $V_{\infty}=3000 \, \rm km \, s^{-1}$, $P_{\rm h}/k_B = 8 \cdot 10^4 \, \rm K \, cm^{-3}$ where a distance of 3.9 Mpc and an age of 250 Myr typical of SBGs have been considered. 
It is possible to observe that the gamma-ray flux (thick red line) as well as the neutrino emission (blue dot-dashed and orange dot-dot-dashed lines) are dominated by the pp interaction taking place in the downstream region, whereas the $p\gamma$ neutrinos are negligible compared to the pp counterpart.
The fluxes of $\gamma$ and $\nu$ are also compared with the CTA Southern Array~\cite{CTA} and IceCube-Gen2~\cite{IceCube-Gen2} sensitivities. 
While a gamma-ray detection in the TeV range of nearby sources seems to be accessible, the detection of local extragalactic wind bubbles by upcoming neutrino observatories appears to be still challenging even though the gap between model predictions and sensitivity will be reduced to less than an order of magnitude.
Adopting the same prototype properties of the SB-wind bubble described above, one can compute the ($p$,$\gamma$,$\nu_{\mu}$) diffuse flux following the star-formation-rate-function approach \cite{Peretti-2,Peretti-wind}.
The multimessenger flux is shown in the right panel of Fig.~\ref{Fig: Multimessenger-Diff} and it is compared with observations by Fermi-LAT \cite{Fermi-LAT<820}, IceCube \cite{IceCube2020_LAST}, IceTop \cite{IceTop2019}, Tunka \cite{Tunka2013,Tunka2016} and Kascade-Grande \cite{Kascade-GRANDE2017}. 
While the total gamma-ray flux from SB-driven wind bubbles contributes approximately to $\sim 5-10 \%$ of the Fermi-LAT data, leaving enough room for SBNi \cite{Peretti-2,Roth-Nature,Ambrosone2020,Owen_diff1} and AGNi \cite{Liu2018,Lamastra_3}, the total neutrino flux and the associated CR flux could provide a sizeable contribution and possibly saturate part of the observed spectra \cite{Peretti-wind,Zhang+Murase2020}. 

A comment on the diffuse flux is in order: most of the calculations in the literature as well as the one presented in this work are performed under the assumption of a prototype source or simple scaling laws. 
Clearly the zoo of all possible sources cannot be reduced to a single prototype. In fact, it is likely that, similar to galaxies, also wind bubbles are distributed on a sequence where the least powerful are dominating in number, whereas the most powerful, able to inject CRs up to EeV, are in the tail of the distribution.
The existence of a luminosity function for galactic wind bubbles would result in a broader and possibly higher CR spectrum at the Earth with a similar result also on the associated gamma-ray and neutrino flux.

To conclude, a comprehensive study of galactic wind bubbles as a population could provide new and insightful clues on the origin of the ($p$,$\gamma$,$\nu_{\mu}$) multimessenger flux at the highest energies and upcoming facilities such as ASTRI Mini-Array \cite{Astri}, CTA \cite{CTA} IceCube-Gen2 \cite{IceCube-Gen2} and KM3NeT \cite{KM3NeT} will guide us with new and exciting observational campaigns.

\section*{Acknowledgments}

This project has received funding from the European Union’s Horizon 2020 research and innovation program under the Marie Sklodowska-Curie grant agreement No. 847523 ‘INTERACTIONS’ and from the Villum Fonden (project n. 18994).

\section*{References}


\end{document}